\documentclass[reprint, amsmath, amssymb, aps]{revtex4-1}

\usepackage{times}
\usepackage{graphicx}
\usepackage{epstopdf}
\usepackage{psfrag}
\usepackage{ae}
\usepackage{amsmath,amssymb}
\usepackage{float}
\usepackage[usenames]{color} 
\usepackage[dvipsnames]{xcolor}
\usepackage[normalem]{ulem}
\usepackage{bm}
\newcommand{\phidn}{$\phi_{\scriptscriptstyle\rm dn}~$}

\begin{document}
\title{Tuning collective actuation of active solids by optimizing activity localization}

\author{Davi Lazzari\textsuperscript{1}}
\email{davi.lazzari@ufrgs.br}

\author{Olivier Dauchot\textsuperscript{2}}
\email{olivier.dauchot@espci.fr}

\author{Carolina Brito\textsuperscript{1}}
\email{carolina.brito@ufrgs.br}

\affiliation{\textsuperscript{1} Instituto de Física, Universidade Federal do Rio Grande do Sul, Caixa Postal 15051, CEP 91501-970, Porto Alegre, Rio Grande do Sul, Brazil}
\affiliation{\textsuperscript{2} Gulliver Lab, UMR CNRS 7083, ESPCI Paris, PSL Research University, 75005 Paris, France}

\date{\today}

\begin{abstract}
Active solids, more specifically elastic lattices embedded with polar active units, exhibit collective actuation when the elasto-active feedback, generically present in such systems, exceeds some critical value. The dynamics then condensates on a small fraction of the  vibrational modes, the selection of which obeys non trivial rules rooted in the nonlinear part of the dynamics. So far the complexity of the selection mechanism has limited the design of specific actuation. Here we investigate numerically how, localizing the activity on a fraction of modes, one can select non-trivial collective actuation. We perform numerical simulations of an agent based model on triangular and disordered lattices and vary the concentration and the localization of the active agents on the lattices nodes. Both contribute to the distribution of the elastic energy across the modes. We then introduce an algorithm, which, for a given fraction of active nodes, evolves the localization of the activity in such a way that the energy distribution on a few targeted modes is maximized -- or minimized. We illustrate on a specific targeted actuation, how the algorithm performs as compared to manually chosen localization of the activity. While, in the case of the ordered lattice, a well educated guess performs better than the algorithm, the latter outperform the manual trials in the case of the disordered lattice. Finally, the analysis of the results in the case of the ordered lattice leads us to introduce a design principle based on a measure of the susceptibility of the modes to be activated along certain activation paths.
\end{abstract}

\maketitle

\section{Introduction:}

A central goal of meta-material design is to realize multi-functionality, enabling a system to effectively perform a variety of tasks. Active solids, composed of elastically coupled active units that locally exert active forces, while being confined to the vicinity of a well-defined reference positions, emerge as promising candidates for achieving such a goal.

Correlated noise generated by an active bath is known to actuate  nontrivial zero modes while suppressing harmonic modes to a degree dependent on temporal correlations~\cite{Woodhouse2018}. This is the simplest evidence for the breakdown of equipartition in active solids.
Active agents embedded in an elastic structure are further able to mobilize solid body motion~\cite{ferrante2013elasticity} or a free-moving mechanism even in a topologically complex case~\cite{Woodhouse2018}. 
Subsequently, experimental and numerical evidence presented in~\cite{baconnier2022selective} revealed that the generic presence of a nonlinear elasto-active feedback of the elastic stress on the orientation of the active forces can induce selective and collective actuation of the solid: a collective oscillation of the lattice nodes around their equilibrium position emerges. Only a few elastic modes are actuated and crucially, they are not necessarily the lowest energy ones.

In the presence of several actuatable zero modes, whether trivially associated with solid body motion or more complex mechanisms, several dynamics coexist in phase space and a general formalism to describe the statistical evolution of collective motion has been derived~\cite{hernandez2024model}. Such coexistence can also hold in the case of mechanically stable solids, as illustrated experimentally by realizing a hysteretic tension-controlled switch between two actuation dynamics~\cite{baconnier2023discontinuous}. Altogether active solids indeed offer a promising horizon for the design of multi-functional meta-materials. Furthermore, most of the active solids considered so far are ordered and hold a spatially homogeneous distribution of active forces, leaving room for a large potential of alternative actuation strategies.

Exploring such an opportunity is the main goal of the present work. To do so, we shall control the injection of energy in the lattice by taming the spatial distribution of activity in both ordered and disordered elastic lattices. More specifically we aim at answering the following question. To what extent, and with which guiding principles, can one design the spatial distribution of activity in the lattice, in order to activate some specific modes? 

We perform numerical simulations of an agent based model on triangular and disordered lattices and vary the concentration and the localization of the active agents on the lattices nodes. We first show that, in sharp contrast with equilibrium solids, the distribution of energy in the elastic modes is very far from being equally distributed and can be controlled by the distribution of activity in the lattice. We then introduce an algorithm, which, for a given fraction of active nodes, evolves the localization of the activity in such a way that the energy distribution on a few targeted modes is maximized -- or minimized. We illustrate on a specific targeted actuation, how the algorithm performs as compared to manually chosen localization of the activity. While, in the case of the ordered lattice, a well educated guess performs better than the algorithm, the latter outperform the manual trials in the case of the disordered lattice. Finally, the analysis of the results in the case of the ordered lattice leads us to introduce a simple design principle based on a measure of the susceptibility of the modes to be activated along certain activation paths.

The paper is organized as followed. In section II, we introduce the agent models, together with the observables, we will used throughout the paper. Section III is devoted to the characterization of the energy distribution among the modes for ordered and disordered networks of different sizes, varying the concentration of active nodes. Section IV explore the selection of actuated modes by tuning the distribution of active particles in a triangular lattice and leads us to propose an optimization algorithm in section V, the performance of which is compared to the "manual" design, in the case of the ordered and disordered lattices. While the algorithm outperform the manual trials in the case of the disordered lattice, a well educated guess performs better in the case of the ordered lattice, suggesting the existence of a simple design rule, which we propose in section VI, before concluding.

\section{Models and Methods}
\label{modmet}

\subsection{Lattices}

\begin{figure}
\centering
\includegraphics[width=1\columnwidth]{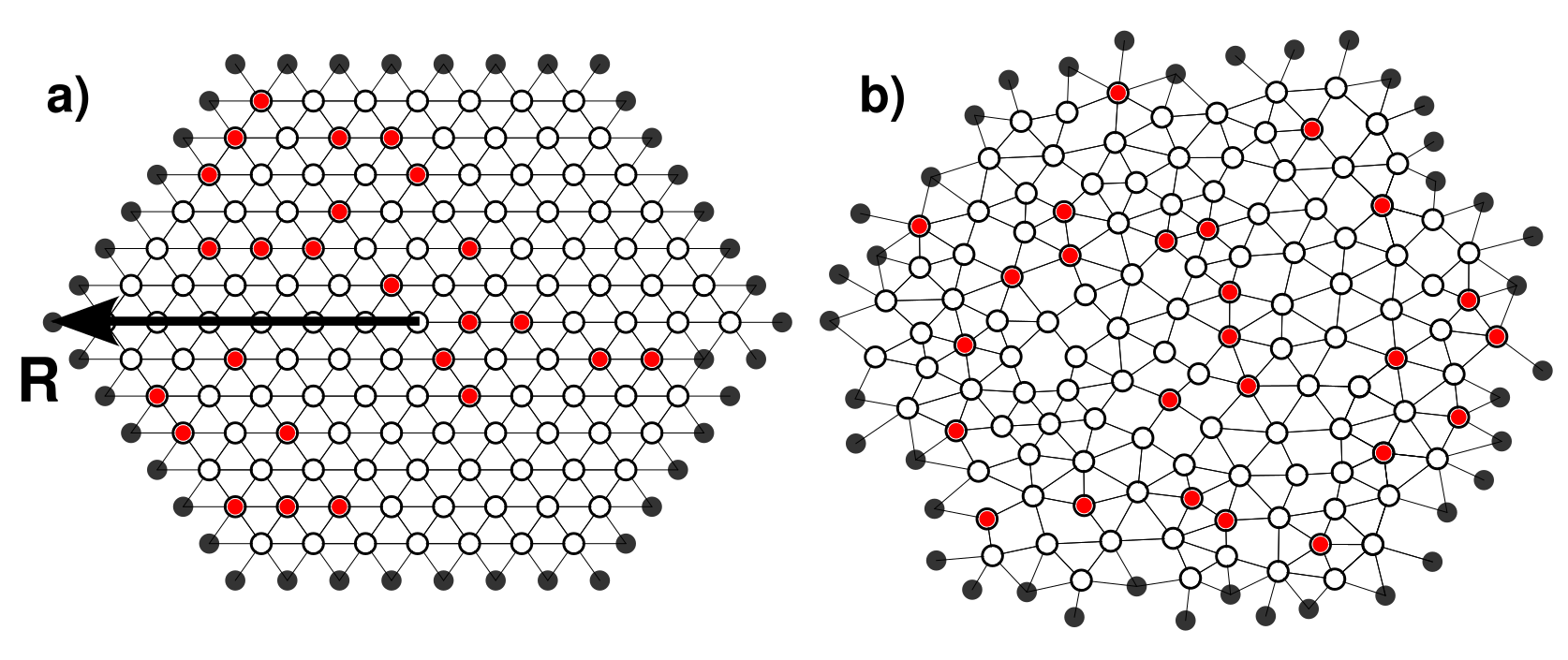}  
\caption{{\bf Ordered and disordered lattices with randomly distributed driven nodes:}  \textbf{(a)} triangular lattice with $N=127$ ($R = 7$) and 25 randomly distributed  driven nodes, resulting in \phidn $= 0.2$.  \textbf  {(b)} a disordered lattice with $N=126$ and \phidn $= 0.2$. In both cases, the external layer of nodes is pinned, as indicated by dark grey points in the figure. White points represent empty nodes and red points thermally or activity driven nodes.}
\label{fig_networks}
\end{figure}

We consider two-dimensional elastic lattices at mechanical equilibrium, consisting of nodes with a well-defined reference configuration, connected by springs of stiffness $\kappa$, Fig. (\ref{fig_networks}). The extremal nodes of the lattice are pinned to the lab frame. Both ordered and disordered lattices are considered. The ordered lattices are triangular lattices, composed of 
$N = 1+3R(R-1)$ nodes, located on $R$ concentric hexagonal rings (see Fig. \ref{fig_networks}-a for an example of such lattice with $R=7$). The disordered lattices are created by first generating a packing of soft discs at high density, following the protocol introduced in~\cite{Brito_PRX2018, LernerPRE2019} (see Appendix).  The center of each disc of the so obtained packing defines a node and two nodes $i,j$ are connected by a spring whenever $r_{ij} < R_j+R_i$, where $r_{ij}$ is the distance between nodes $i$ and $j$ and $R_j$ is the radius of the disc $j$. 
To compare ordered and disordered lattices of the same sizes, large disordered lattices are generated, inside which we pin an hexagon of nodes at a distance $d \approx R+1$ from the center of the system in such a way to have approximately $N(R)$ moving nodes within the pinned boundary (see Fig. \ref{fig_networks}-b) for an example of such lattice with $R=7$.
All disordered networks used throughout the work have high coordination number ($z \approx 6$) and their density of states $D(\omega_k)$ -- defined as the distribution of the frequencies $\omega_k$ -- are statistically similar, for low frequencies, to the ordered case, as shown in Fig.~\ref{Dw_ord_dis} of the Appendix \ref{sec_appendix_disordered_lattice}.

\subsection{Dynamics}

It was shown that the phenomenology of collective actuation is very well captured within the harmonic approximation, where the elastic forces are entirely encoded into a linear description $\bm{F}^{el}_i = -\mathbb{M}_{ij} \bm{u}_j$, with $\bm{u}_i$ the displacement of node $i$, and $\mathbb{M}_{ij}$ the dynamical matrix of the elastic lattice of interest. Here we follow the same scheme, but, in contrast with previous works, only a fraction $\phi_{dn}=n_{dn}/N$ of nodes are driven. We mostly consider active driving, but also introduce thermally driven network for comparison.

In the active case, the driven nodes obey the self-aligning overdamped dynamics introduced in~\cite{baconnier2022selective}.
Each driven node is activated by a free to rotate self-aligning active particle that exerts a polar force on the node, while being reoriented by the total elastic force acting on that node. This nonlinear elasto-active feedback is the key ingredient responsible for the onset of collective actuation (see also~\cite{Baconnier_review2024} for a review on self-aligning polar particles). The dynamical equations read:  	
\begin{eqnarray}
    \bm{\dot u}_i &=& \Pi \bm{\hat n}_i - \mathbb{M}_{ij} \bm{u}_j,
    \label{eq_active1}
    \\
    \bm{\dot n}_i &=& \bm{\hat n}_i \times \bm{\dot u}_i \times \bm{\hat n}_i,
    \label{eq_active2}
\end{eqnarray}
where the unit vector $\bm{\hat n}_i$ indicates the direction in which the particle sitting at node $i$ exerts the active force. $\Pi~=~l_e / l_a$, the unique control parameter is the ratio of two length, the elastic length $l_e=F_a /\kappa$, which describes the elongation of a spring of stiffness $\kappa $ under the action of an active force $F_a$ and the alignment length $l_a$, which is the typical length a node $i$ must be displaced to reorient the active particle sitting at this node. Eq.~(\ref{eq_active2}) describes the reorientation of particle $i$ towards its displacement $\bm{\dot u}_i$ according to the self alignment mechanism. A noise term can be added to Eq.~(\ref{eq_active2}), we discuss its role in Appendix \ref{noise}. In this active setting, the non driven nodes simply obey the same equation, with zero activity, that is $\Pi = 0$.

In the thermal case, {it was chosen} the driven nodes {to} obey the standard {underdamped} Langevin equation{, since it estimates the thermodynamics properties better than the overdamped case},
\begin{eqnarray}
    \bm{\Ddot{u}}_i = - \bm{\dot u}_i - \mathbb{M}_{ij} \bm{u}_j + \sqrt{2 T_{\rm eff}} \bm{\xi}_i(t),
    \label{eqLang}
\end{eqnarray}
where $\bm{\xi}_i(t)$ is a Gaussian white noise with zero mean $\langle \bm{\xi}_l(t) \rangle = 0$, $\langle \bm{\xi}_l(t) \cdot \bm{\xi}_k(t') \rangle = \delta(t-t')\delta_{lk}$. $T_{\rm eff}$ controls the noise amplitude. The non driven modes simply respond elastically and transfer the elastic forces around the network obeying newtons law.

\textbf{Numerics:} In the case of active driving the dynamical equations were integrated using a Runge-Kutta 4th order method with $dt = 0.01$. For the thermal driving, the stochastic equations were integrated using the Stochastic Velocity Verlet algorithm \cite{gronbech2020complete} (also $dt = 0.01$). The measures of interest are taken on the interval $t \in [250,500]$ time steps ($\Delta t = 250$), disregarding the transient regime.

\subsection{Observable}
\label{dm}

We are primarily interested in the distribution of the elastic energy among the vibrational modes of the lattices. These modes --  also called normal modes (NM) -- are the eigenvectors $|\bm{\varphi}_k \rangle$ of the $2N \times 2N$ symmetric dynamical matrix and form a  complete orthonormal basis. The lattices considered here being all mechanically stable, they are associated with strictly positive eigenvalues $\omega_k^2$.  In the case of the triangular lattice, it is convenient to sort these modes by the four classes of rotational symmetries, which we will simply denote symmetry of class 1, 2, 3, 4, that leave the lattice and its normal modes invariant.  Examples of the normal modes classified by their classes of symmetries are shown in the appendix~\ref{symmetries} for a system with $R=7$ (see also Supp. Matt in~\cite{baconnier2022selective} for an explicit construction).  The dynamics is then decomposed on these modes, by projecting the $2N$-dimensional displacement field $| \bm{u} \rangle =  \{\bm{u}_1, \bm{u}_2, .., \bm{u}_N\}$ on each mode: $P_k(t) = \langle \bm{\varphi}_k | \bm{u}(t) \rangle$. 
In practice, $P_k(t)$ is averaged over time in the steady state and normalized as follows:
\begin{eqnarray}
     \overline{P_k^2} = \frac{\overline{\langle \bm{\varphi}_k | \bm{u} \rangle^2 }}{\sum_k \overline{\langle \bm{\varphi}_k | \bm{u} \rangle}^2},
     \label{Pk2medio}
\end{eqnarray}
where $\overline{x}$ denotes a temporal average over 250 time steps (disregarding the first 250 steps).

\section{Energy distribution for varying fraction of driven nodes}
\label{sec_varying_concentration_ap}

\begin{figure}[t]
\includegraphics[width=0.9\columnwidth]{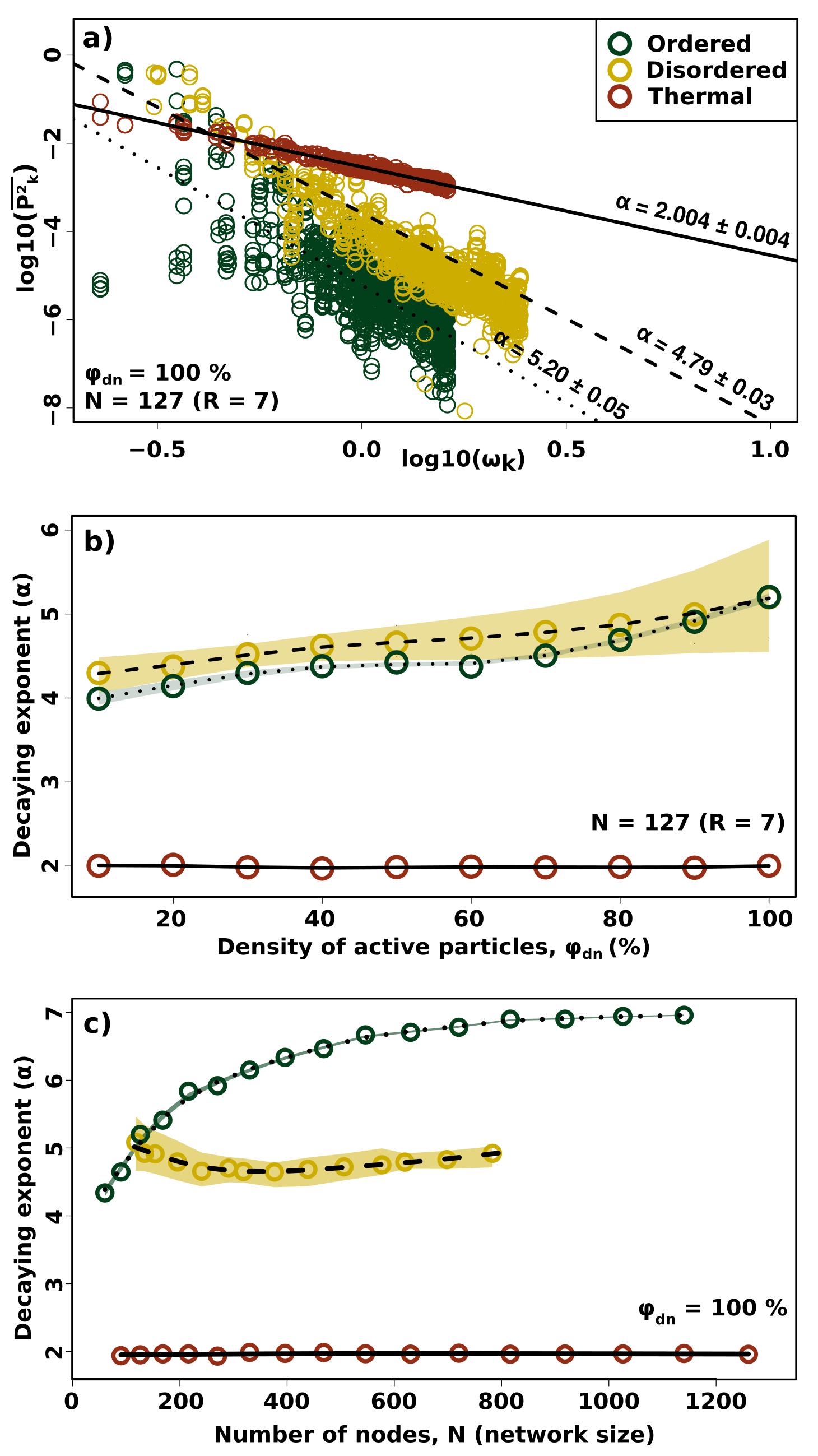}
\caption{{\bf Distribution of energy in thermally and actively driven lattices:} {\bf (a)} Average squared amplitude of the modes $\overline{P_k^2} \propto \omega_k ^{\alpha}$ {\it vs} $\omega_k$ for a thermally driven, an actively driven ordered, and an actively driven disordered lattice as reported in the legend.  Each point corresponds to one mode. The dashed line corresponds to the fit $\overline{P_k^2} \propto \omega_k ^{-\alpha}$. While $\alpha=2$ in the thermal case, it is much larger in the active case. $N=127, \phi_{dn}=100\%$.  {\bf (b)} Dependence of $\alpha$ on the fraction of driven nodes $\phi_{dn}$ for $N=127$. {\bf (c)} Dependence of $\alpha$ on the system size $N$ for $\phi_{dn}=100\%$. The data are averaged over 50 initial conditions and 12 different disordered lattices where used. ($\Pi = 1.3$).} 
\label{TermicovsAtivo}
\end{figure}

We first examine the distribution of energy among the modes when the driven nodes are randomly distributed in the lattice. We consider both disordered and ordered lattices and vary both the fraction of driven nodes $\phi_{dn}$ and the total number of nodes $N$. At first order, the elastic energy of the system can be written in terms of the matrix $\mathbb{M}$:
\begin{eqnarray}
    \Delta U = \langle \bm{u} | \mathbb{M} | \bm{u} \rangle = \sum_{k=1}^{2N} \omega_k^2 P_k^2.
    \label{eq_potential_energy}
\end{eqnarray}
At equilibrium, equipartition dictates that each quadratic degree of freedom contributes equally to the system's energy, typically $k_BT/2$ per degree of freedom, where $k_B$ is the Boltzmann constant. In the present context, this imposes that each term of the above decomposition contributes an equivalent amount of energy, so that $P_k^2 \propto \omega_k^{-2}$. This is precisely what is reported on Fig.~\ref{TermicovsAtivo}-a for the thermally driven case, when all nodes are driven.

\begin{figure*}[t!]
\centering
\includegraphics[width=1.999\columnwidth]{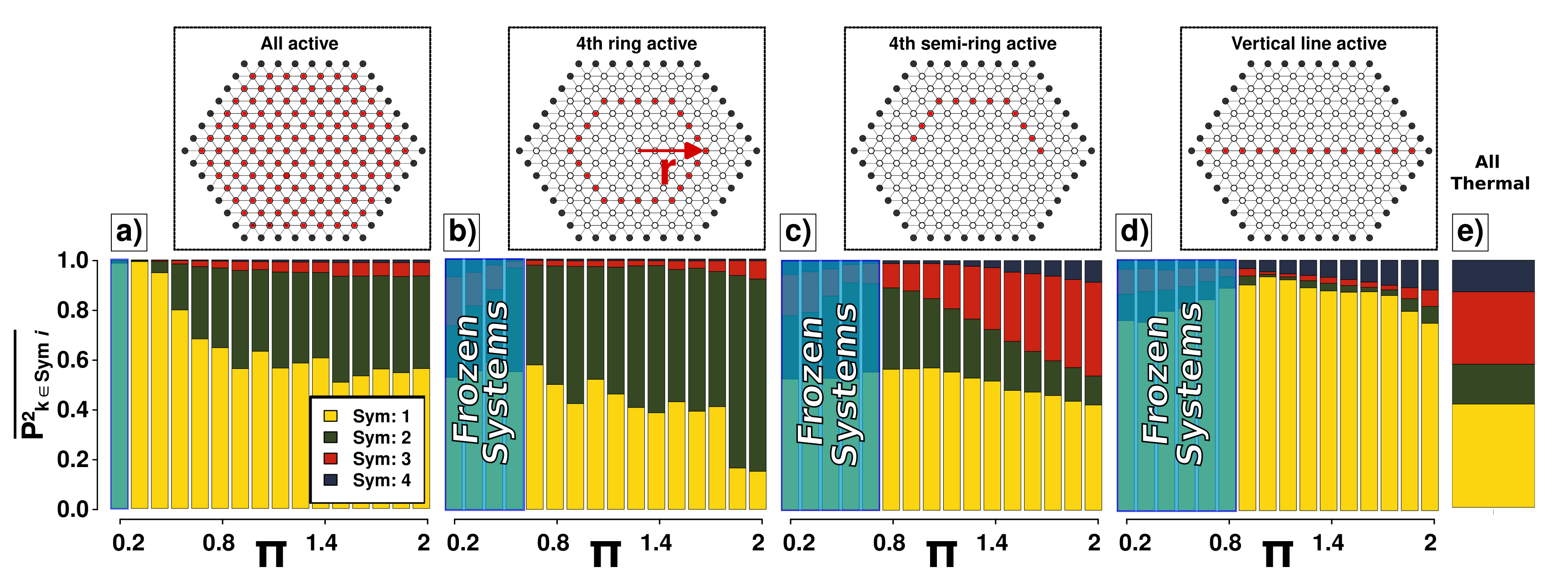}  
\caption{{\bf Selective actuation of modes by symmetry classes, using different spatial distribution of active nodes: } The top row indicate the localization of the active nodes. Black points represent pinned edges, red points active nodes and open circles are empty nodes. In {\bf (a)} active particles are located in all the nodes of the lattice, {\bf (b)} in the 4th ring,  {\bf (c)} in a semi-circle of the 4th ring and  {\bf (d)} in an horizontal line.  The bottom row indicates the energy repartition among the modes grouped by symmetry classes, as indicated by the colors of the legend. The extreme right column {\bf (e)} indicates this distribution when all nodes are thermally driven, as a reference. For each spatial distribution (column (a), (b), (c) (d)), the distribution of the energy is represented for increasing values of $\Pi$. When $\Pi$ is too small, collective actuation does not take place and the system remains frozen, as indicated by the blue overlay. Data shown here are averaged over 100 runs with random initial polarization;  lattice size $N=127$.}
\label{fig_projection_symmetries}
\end{figure*}

In the case of active driving, equipartition does not hold. We start by simulating lattices of size $N=127$, for which all nodes are actively driven ($\phi_{dn}=100\%$). Quite remarkably, one still observes a power-law dependence $P_k^2 \propto \omega_k^{-\alpha}$, for large enough $\omega_k$, albeit with a much larger value of $\alpha$, indicating a stronger condensation of the energy on the low energy modes, the effect being more pronounced in the case of ordered lattices.
A strong condensation in the softest modes were also observed in the context of jammed active system \cite{henkes2011active}.
These results are confirmed, whenever varying the fraction of driven nodes (Fig.~\ref{TermicovsAtivo}-b), or the lattice size (Fig.~\ref{TermicovsAtivo}-c).  The exponent $\alpha$ is very robust with respect to the fraction of driven nodes. Even for fraction as low as $\phi_{dn}=10\%$, $\alpha\simeq 4$ remains twice larger than its equilibrium counterpart. We also note that in the case of ordered lattice, increasing the system size amplifies the condensation, with $\alpha$ reaching values as large as close to $7$, where it seemingly saturates. This is not the case for disordered lattices, where $\alpha\simeq 4.5$ for all system sizes probed here. In all cases, the violation of equipartition reported above opens the path for manipulating the energy injection, in view of actuating modes preferentially. 

\section{Modes selection by tuning the spatial distribution of active nodes}

We first concentrate on the ordered lattice and investigate how the spatial distribution of the actively driven nodes condition the energy distribution amongst the modes, the modes being grouped according to their four classes of symmetries (see Appendix \ref{symmetries}).  We consider four distinct spatial organization of the actively driven modes, corresponding to the four columns of Fig.~\ref{fig_projection_symmetries}:  (a) all active, (b) the 4th ring only is active, (c) half of the 4th ring is active, (d) a centerline is active. These choice are somehow arbitrary but illustrate well the role of the spatial localization of the active nodes. The results are shown for increasing values of the activity, as indicated by the dimensionless parameter $\Pi$.

As a first observation, we note that the active driving leads to very different distributions of the elastic energy across the modes of different symmetry classes. When all nodes are driven (a), the modes of class 1 and 2 vastly dominate the dynamics, as compared to the passive case, where the equipartition of energy favors the modes of class 1 and 3. Second, one sees that the distribution of energy is only slightly changed when reducing the activation to a unique line on the fourth ring (b). Conversely the distribution of energy among the four classes of symmetry is strongly altered when only half of the ring is driven (c), or when only the central line is driven (d) pointing at an important role of the symmetry of the spatial distribution of the driven nodes. We also note that, depending on the geometrical organization of the driven nodes, the dependence on $\Pi$ can be minimal, as in case (a,b,d), or pretty strong as in case (c). 

These observations drive us to search for spatial distributions of the active nodes that enhance the projection of the dynamics in a {\it desired} class of symmetry or, even, in a few specific number of modes.

\section{Optimization algorithm}

Can one determine a spatial distribution of active particles that effectively amplifies the dynamics within a desired class of symmetry or a specific normal mode? To address this question, we propose an algorithm that combines molecular dynamics simulations of the dynamics and the Metropolis Monte Carlo method to evolve the active configurations of the lattice.

A spatial distribution of actively driven nodes is denoted by a vector $|\sigma\rangle = \{0,1,1,0,\cdots\}$ of size $N$, where the one's indicate the nodes that are actively driven, and the zero's the non driven nodes. 
\begin{figure}
\centering
\includegraphics[width=0.9\columnwidth]{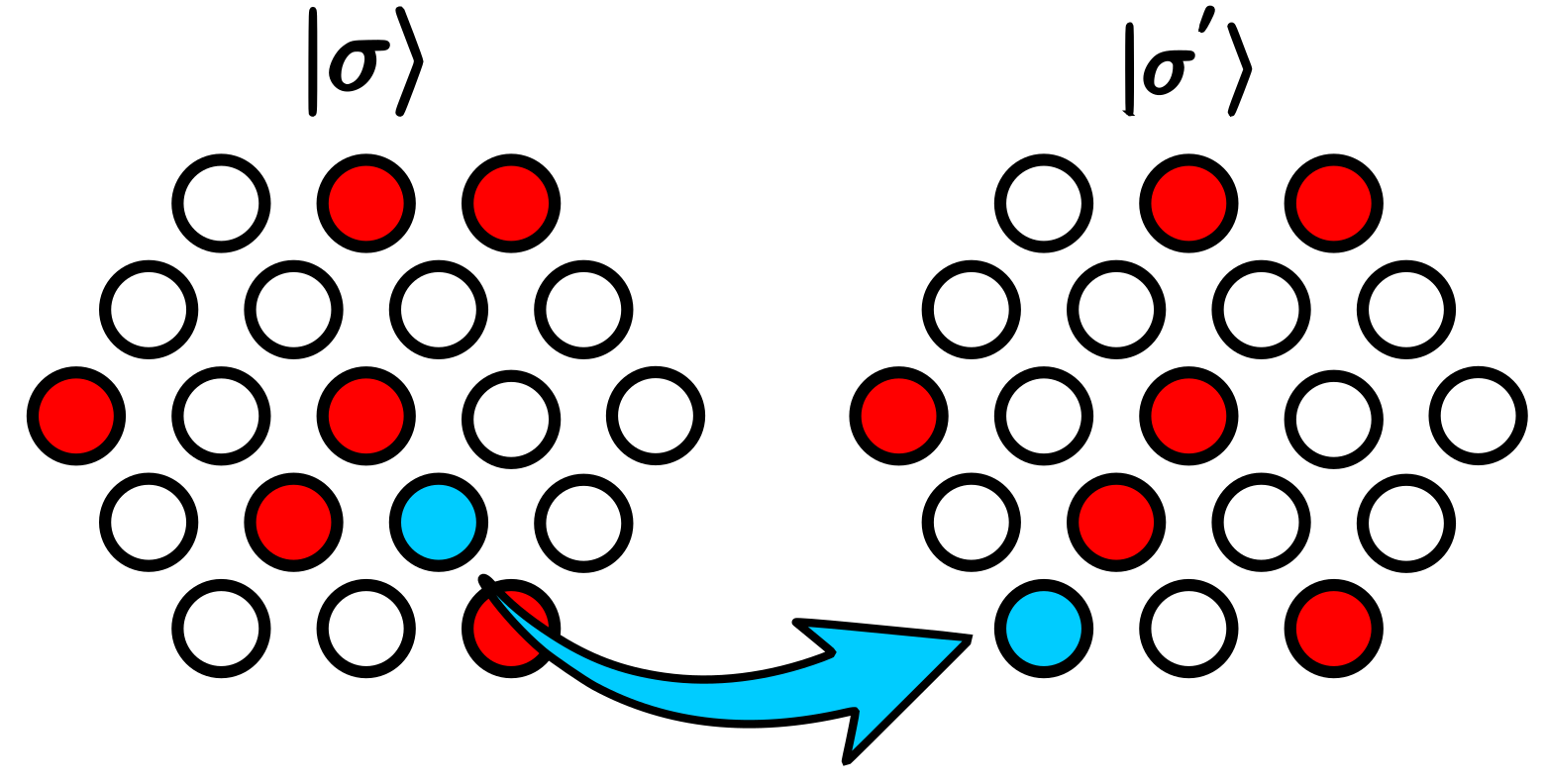}  
\caption{{\bf Monte Carlo move:} schematic example of two configurations, $|\sigma\rangle$ and $|\sigma^{'}\rangle$, which differ by one Monte Carlo step, where an active node is made inactive in favor of an other node, keeping the fraction $\phi_{dn}$ constant.}
\label{esquema}
\end{figure}
\begin{figure*}
\centering 
\includegraphics[width=1.8\columnwidth]{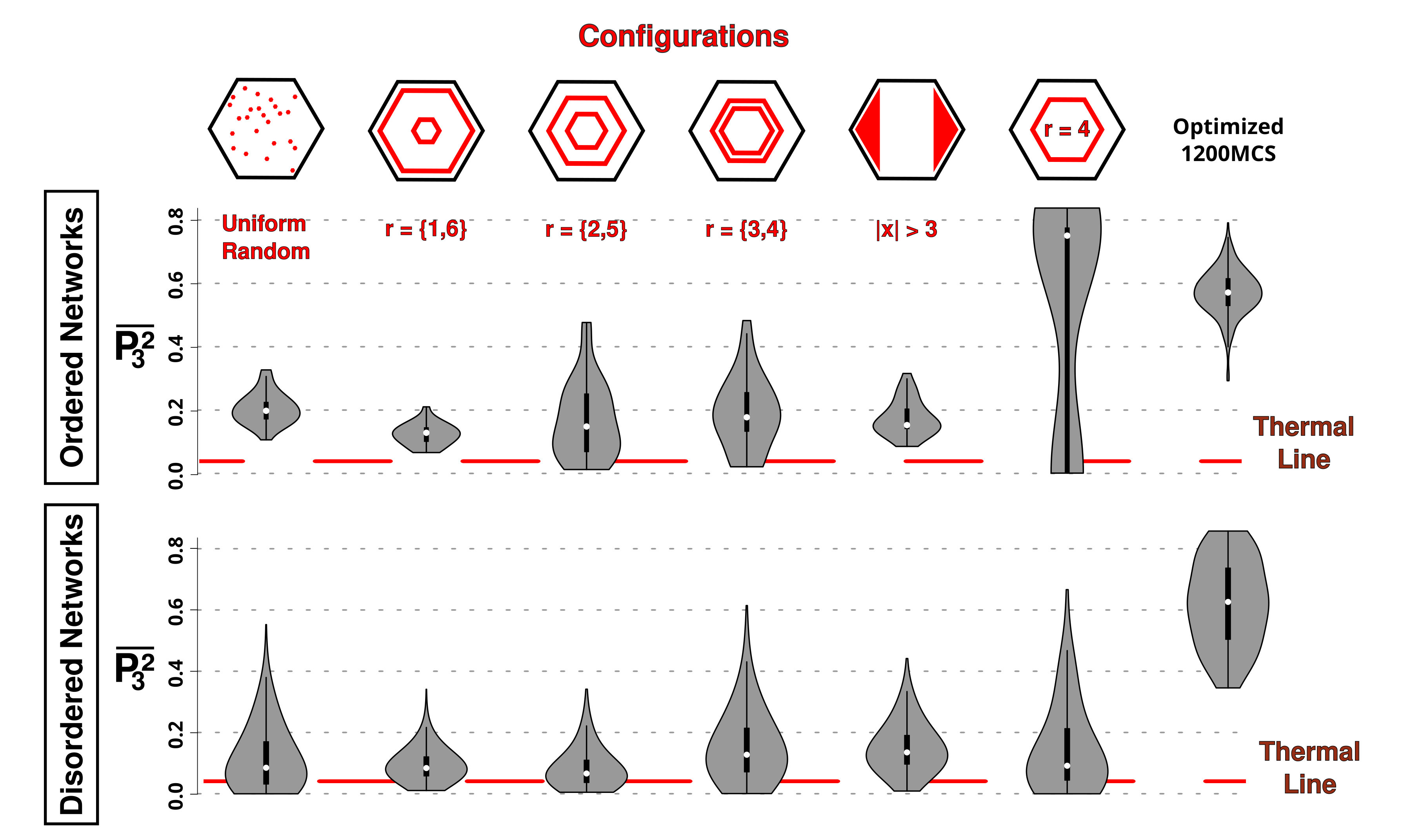}    
\caption{{\bf Performance of the optimization algorithm:} Violin plot showing the distributions of the final projections onto mode 3, denoted  $C_3^{|\sigma\rangle} = \overline{P_3^2}$, obtained after 1200 Monte Carlo steps (last column on the right). These distributions are compared across 6 preset configurations with spatial distributions of the active nodes, indicated in red on the top row. The white points represent the mean of the distribution, while the black bars indicate the dispersion around the mean. The density of active nodes,  $\phi_{dn} = 24/127$ is identical across all cases. The results are shown for both ordered hexagonal lattices (top) and disordered lattices (bottom). In the ordered networks, 30 distinct initial conditions (ICs) were used for each configuration, except for the optimization case, for which were used 350 distinct ICs. In the disordered networks, 30 ICs were simulated for 30 different networks in all configurations, except for the optimized cases, which involved only 10 networks. $R=7$, $N=127$ and $\Pi = 1.3$.}
\label{fig_opt}
\end{figure*}

This configuration is evaluated by running the dynamics during $2\Delta t = 200$ time steps and computing the cost function $C_k^{|\sigma\rangle} = \overline{P_k^2} [\sigma]$, where the temporal averaged runs over the $\Delta t = 100$ time steps composing the second half of the simulation time window (the choice of this time window is discussed in the Appendix \ref{sec_time_window}). $C_k^{|\sigma\rangle}$ evaluates the ability of the configuration $|\sigma\rangle$ to concentrate the energy of the system in a given mode $k$. Every $2\Delta t$ time steps, a Monte-Carlo move is proposed from the configuration $|\sigma\rangle$ to another configuration $|\sigma'\rangle$, by changing the location of one active node, hence keeping the overall fraction of active nodes constant (see Fig.~\ref{esquema}).  
The goal being to maximize, respectively minimize, the cost function, depending on wether one wants to increase, respectively decrease, the projection on a given mode, the new configuration $|\sigma^{'} \rangle$ is accepted with probability:
\begin{eqnarray}
	P(|\sigma \rangle  \rightarrow  |\sigma^{'} \rangle) = 
 	\min \left\{1, \exp\left(-\frac{C_k^{|\sigma^{'}\rangle}- C_k^{|\sigma\rangle}}{T_{e}}\right)\right\},
 \label{eq_prob_change}
 \end{eqnarray}
where $T_e$ is an effective temperature which allows for the exploration of the configurations space. If the new configuration $|\sigma^{'}\rangle$ is accepted, the next step starts from it; if not, the original configuration $|\sigma\rangle$ is restored and another move is proposed. A Monte Carlo Step (MCS) is defined as $N$ trials configurations. We use $T_e=10^{-3}$ and test the dependence of our results on $T_e$ in the Appendix \ref{det}, Fig. \ref{fig_variaTe}. 

\begin{figure*}
\centering
\includegraphics[width=1.8\columnwidth]{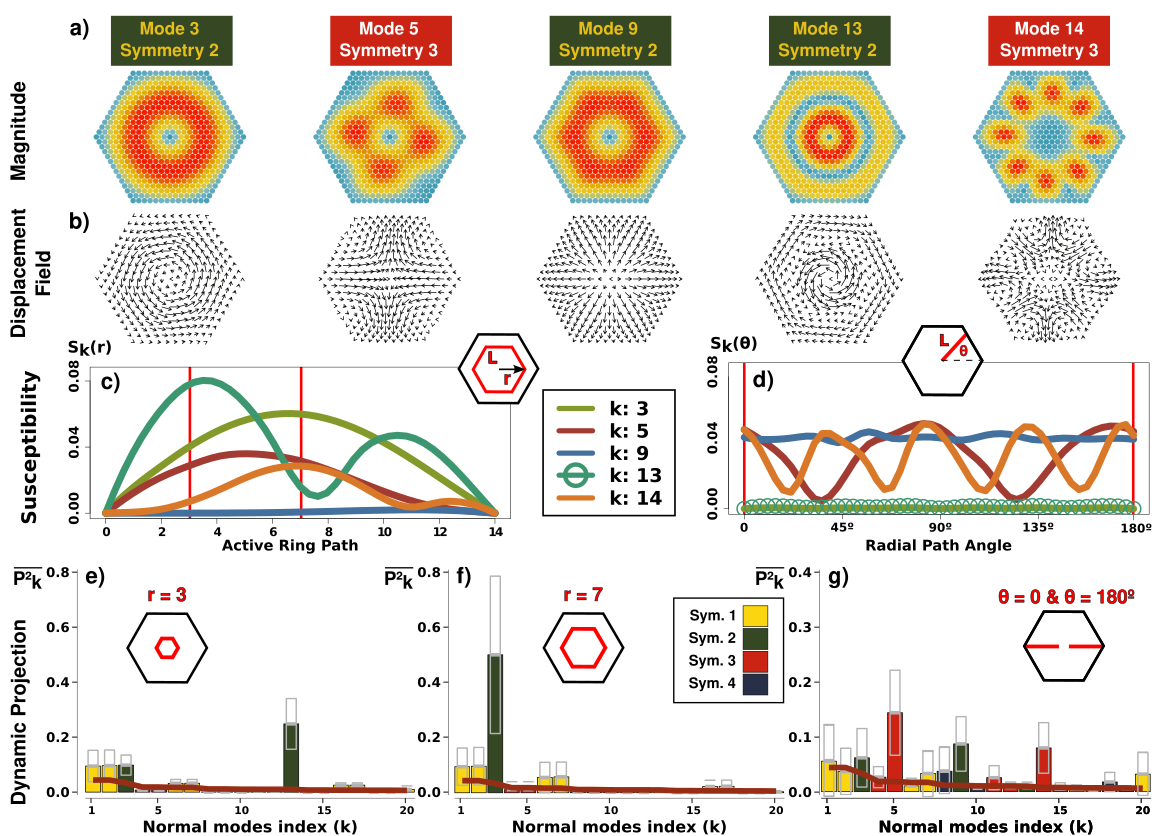}  
\caption{{\bf Optimization of the selective actuation of hexagonal lattices:}  {\bf (a,b)}  Five specific modes of the triangular lattice, of size $N=547$, $R=14$,  ($k=3,5,9,13,14$), with (a) the amplitude and (b) the vector field of the local polarization of the mode, (red, respectively blue, regions indicate large, respectively small, displacement in the mode polarization.  
 {\bf (c)} The susceptibility $S_{k}(r)$ as a function of $r$ and {\bf (d)} $S_{k}(\theta)$ as a function of $\theta$  for the five selected modes of interest as indicated in the legend. { \bf (e-h)} Respective condensation of the energy $\overline{P_k^2}$ on modes $k$, colored by their class of symmetry, as indicated in the legend of panel (f), for four configurations of active nodes, as indicated in each panel by the sketch of the activation path in red; {\bf (e)}: a ring on $r=3$; {\bf (f)}: a ring on $r=7$; {\bf (g)}: two segments on $\theta=0^{\circ}$ and $\theta=180^{\circ}$; 
 }   
 \label{fig_selection_rules}
\end{figure*}

The algorithm is evaluated on its ability of finding a configuration, for which the spatial distribution of $24$ active nodes among $127$ nodes maximizes the condensation of the dynamics on the mode $k=3$, as indicated by the value of $\overline{P_3^2}$. The mode $k=3$ belongs to the class of symmetry 2 and is therefore not predominantly actuated in a typical configuration. We recall that the mode 3 is shown in the Fig. \ref{fig_selection_rules}-(a,b). 
The performance of the configuration optimized by the algorithm after $1200$ Monte Carlo steps (last column on the right) is compared to configurations, where the same number of active nodes are localized in preset geometries (first six columns) as indicated by the drawings in red, on the top row of Fig.~\ref{fig_opt}: randomly chosen around all the lattice (Uniform Random); localized on two layers $r = \{1,6\}$, $r = \{2,5\}$, and $r = \{3,4\}$; on two separated areas, $|x| > 3$ which do not respect the rotational symmetries of the lattice; with the 4th layer fully occupied, $r = 4$. 
The top, respectively the bottom, row displays the distributions obtained for the ordered, respectively the disordered lattices for 30 different initial conditions. In the case of the disordered network, we generated 30 different lattices -- except for the optimized case where we used 10 -- and, for each one, 30 initial conditions where simulated. 



In general, the algorithm performs much better than the preset configurations. In the disordered lattices this is always the case. In the case of the ordered hexagonal lattice, the activation of the 4th layer, which precisely match the location of maximal polarization inside the mode $k=3$, performs better. Note that the dynamics being deterministic, a fraction of initial condition lead to complete failure of the activation of the third mode, when the spatial distribution of the active nodes is preset, while the optimization algorithm allows for an adaptation of the spatial distribution of the active nodes to the randomly chosen initial condition.

\begin{figure*}
\centering
\includegraphics[width=1.8\columnwidth]{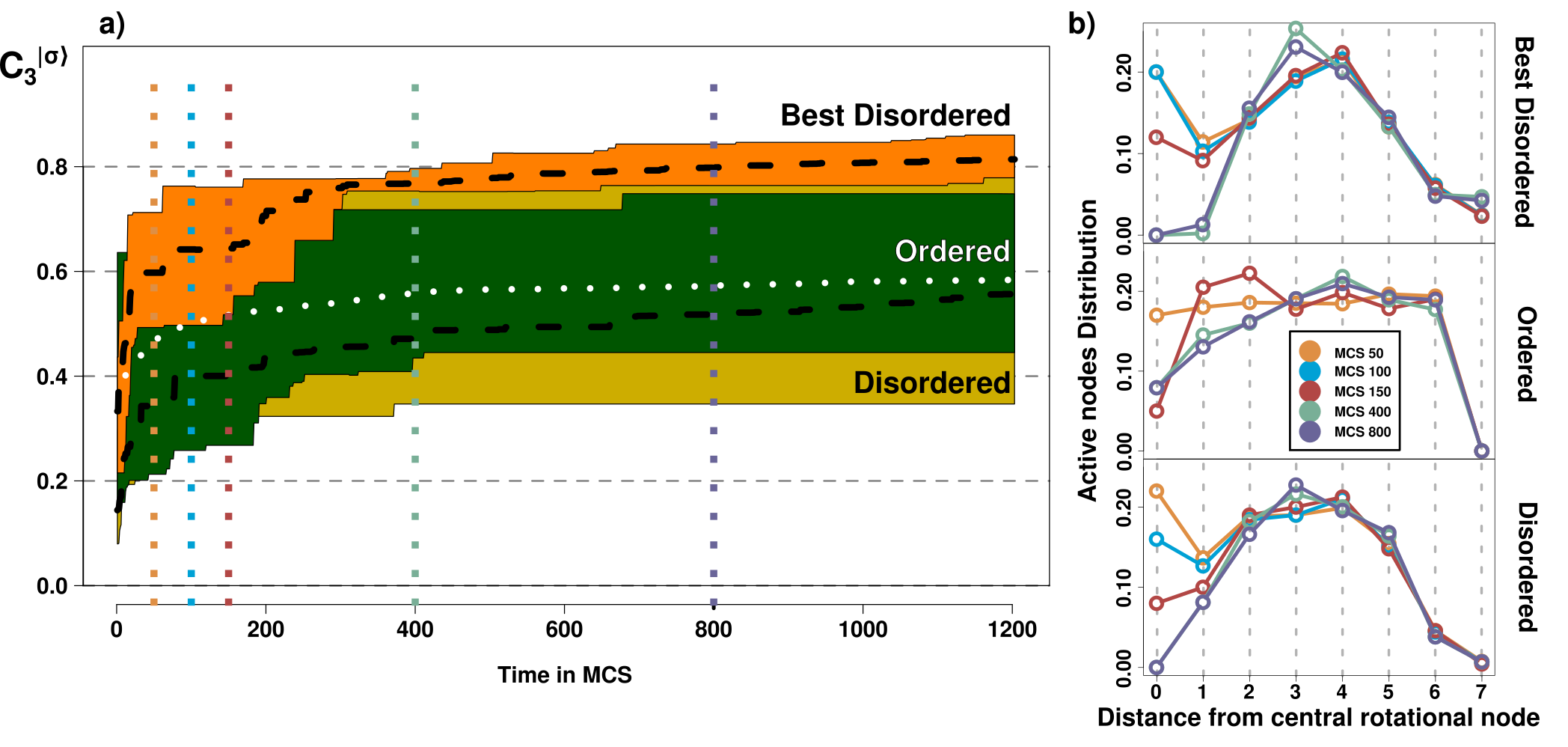}  
\caption{{\bf Convergence of the algorithm performance:} \textbf{(a)} Evolution of the cost function for the projection on the mode $k=3$, $C_3^{|\sigma\rangle} = \overline{P_3^2}$, with the number of Monte Carlo steps in three different networks: the ordered hexagonal lattice (green and white dotted line), a randomly chosen disordered lattice (yellow and black dashed line) and the disordered lattice, for which the algorithm obtains the best performance (orange and black dashed line). The lines indicate the mean performance over $100$ random initial conditions, while the colored areas are delimited by the worst and best cases.
\textbf{(b)} Spatial distribution of the active nodes as a function of their distance to the central one at different moment of the optimization process (as indicated in the legend) for the three networks (as labeled on the right of each panel); ($R=7, N=127$; $\Pi = 1.3$).}
\label{fig_conv_performence}
\end{figure*}

In Figure~\ref{fig_conv_performence}, we present the convergence of the algorithm for the case where the goal is to maximize the projection onto mode 3. We also analyze the distribution of active particles in the lattices as a function of the algorithm's convergence.
Figure~\ref{fig_conv_performence} displays the evolution of the performance of the algorithm as a function of the number of Monte Carlo steps for three different lattices: the ordered hexagonal one, a  typical disordered lattice, and the disordered lattice, for which the algorithm obtains the best performance. For each of them the figure shows the mean performance, averaged over $100$ initial conditions, while the worst and best cases delimitate the colored areas. From these evolutions one sees that disordered networks can achieve a better performance than the ordered one, in the sense that the optimization algorithm identifies a spatial distribution of the active nodes that favors a stronger condensation of the dynamics on the selected mode of interest.
From the dynamical evolution of the localization of the active nodes during the optimization process it appears that the algorithm has a hard time in identifying the best configuration in the case of the ordered lattice. More specifically, it only very slowly condensates the active nodes toward the 4th layer of the lattice, in contrast with the disordered case, where this condensation takes place in the early steps of the optimization. We interpret this behavior as a sign that the symmetries of the ordered lattice favors the nearly degenerescence of many configurations with respect to their evaluation by the cost function.

\section{Optimization rule for the selective actuation of hexagonal lattices}

The analysis of the results from the optimization algorithm suggests a possible simple strategy for optimizing the spatial distribution of the active nodes in order to achieve the condensation of the dynamics on some specific modes. We observed that the optimal configuration to amplify the projection of the dynamics on the third mode of the hexagonal lattice, for a network size with $R=7$, corresponds to a localization of the active nodes on the fourth layer of the lattice. As noted already, this coincides with the regions of largest magnitude of the displacement field in mode $k=3$ (see Appendix~\ref{symmetries}). 

We propose to generalize this observation and organize the spatial distribution of the active nodes in light of the displacements field geometry of the mode of interest. The underlying hypothesis is that localizing the active nodes in the regions of high displacement of a mode should favor the coupling of the activity to that mode. To do so we first define an {\it activation path} $L$ on the lattice, composed of $|L|$ adjacent nodes and define $S_{k}(L)$ the {\it activation susceptibility} of that path in mode $k$ as the local projection of the polarization field on the tangent to this path: 
\begin{eqnarray}
	S_{k}(L) = \sum_{l=0}^{|L|} \frac{| \bm{\hat e}_l \cdot (\bm{\varphi_k})_l| }{|L|},
\end{eqnarray}
where $l$ indexes the nodes along the path $L$, $\bm{\hat{e}}_l$ is the unitary vector tangent to the path at node $l$, and $(\bm{\varphi_k})_l$ is the displacement of mode $k$ at node $l$. This susceptibility is defined such that it is maximal when the path $L$ runs parallel to the local maxima of the displacement field of the mode of interest. The configurations of active lines transport the alignment information along the path back and forward, favoring oscillations to occur parallel to the path.

Figure~\ref{fig_selection_rules} illustrate the application of the above ideas in the case of an ordered hexagonal lattice with $R=14$ layers, for which two types of active paths are tested: (i) concentric active rings of size $r$ ($L(r)$, with $|L(r)| = 6 r$ and $\bm{\hat e}_l = \bm{\hat e}_\theta(l)$, see Fig.~\ref{fig_selection_rules}-c) and (ii) linear radial paths that do not cross the center ($L(\theta)$, with $|L(\theta)| = R-2$ and $\bm{\hat e}_l = \bm{\hat e}_r(l)$, see Fig.~\ref{fig_selection_rules}-d).
Figure~\ref{fig_selection_rules}-(c,d) shows the susceptibility $S_k(r)$, for the concentric ring path of radius $r$ and $S_k(\theta)$ for the radial path with orientation $\theta$ for the five modes $k=3,5,9,13,14$ displayed on the top rows. The core observations is that the susceptibility strongly depends on the combination of the path and the mode. For instance $S_9(r)$ is negligible for all concentric path, while  $S_9(\theta)$ is systematically large for all radial path. The dependence on the path can be relatively simple, as it is the case for $S_3(r)$, where one recovers the observation made earlier that the mode $k=3$ has a maximal susceptibility when the distribution of active nodes concentrate on a ring of radius $r=R/2$. But it can also be less obvious for higher modes with less symmetric displacements, such as mode $k=5$ or $k=14$. 
One also note that, once a path is chosen, there are several modes with susceptibility greater than zero. This implies that these modes would be activated if the nodes were excited along that path. Therefore, optimizing for a single mode is generally likely to be impossible.

Figure~\ref{fig_selection_rules}-(e,f,g) show the actual distribution of energy among the modes for four different paths. For the concentric ring path, one verifies very clearly the strong selection of mode $k=13$ by the ring of radius $r=3$ (Fig.~\ref{fig_selection_rules}-e) and the even stronger one for the mode $k=3$, when active nodes are distributed along the ring of radius $r=7$ (Fig.~\ref{fig_selection_rules}-f).  
The case of the radial paths confirms that the mode $k=9$ is excited for both path configurations. More interestingly one sees that the radial path along $\theta=0^{\circ}, 180^{\circ}$ is unable to select differentially the modes $k=5$ and the mode $k=14$, while the path along $\theta=0^{\circ}, 180^{\circ}$ does select the mode $k=14$, without activating the mode $k=5$.

Altogether, the use of the activation susceptibility is thus a good design principle, although it clearly also reveals the limitation of the selectivity that can be reached. Nevertheless, we observe that the larger the system is, the more selective the activation design can be, especially for modes of high energy, because of the larger specificity of their polarization geometry and the possibility of combining several activation path.

\section{Conclusions} 
In this paper, we explored different strategies for injecting energy into elastic lattices by exciting their nodes with active agents. We investigated both ordered triangular lattices and disordered lattices with a coordination number of approximately $z \approx 6$, similar to that of the ordered lattice. By distributing the active nodes in various spatial organization, we demonstrate the possibility of tuning, at least to some extent, the energy partition amongst the mode. This of course contrast with the classical scenario of equipartition imposed by thermal equilibrium.

When all lattice nodes are active ($\phi_{\text{dn}}=1$), a pronounced concentration of energy is observed on lower-frequency modes, as previously observed both numerically and experimentally~\cite{Woodhouse2018,baconnier2022selective}. More specifically, the amplitude distribution of the energy among the modes exhibits a power-law decay at large enough frequency, $\overline{P_k^2} \propto \omega^{-\alpha}$, with an exponent $\alpha$ significantly larger than 2, the value indicative of equipartition. 
When reducing the fraction of active nodes, the distribution of energy among the modes is heavily influenced by the spatial distribution of the active nodes inside the lattice. While complete control over energy concentration in specific modes remains elusive, we nevertheless demonstrate that there is room for optimization.

In the case of ordered lattice, a well educated guess is possible and we propose a simple design principle, which identifies optimal paths for the distribution of the active nodes, on the basis of the geometry of the spatial structure of the modes, characterizing the path by their activation susceptibility.  
In the case of disordered lattice, specific spatial distribution of the active nodes, combined with specific realization of the disorder, can better condensate the energy on a class of modes of interest, as compared to an ordered lattice with the same size. This suggests that the optimization algorithm has greater potential for evolution toward an "optimal distribution" in the presence of disorder. An interesting perspective would be to jointly optimize for the disorder of the network and the spatial distribution of the active nodes.

When considering disordered lattices, our focus has primarily been on cases where the coordination number is high, approximately $z \approx 6$ and the lattice is largely hyperstatic and mechanically stable. It would be valuable to extend our investigation to lattices with lower coordination numbers, approaching $z \rightarrow 4$. In this limit, it is known that the lattices are on the verge of losing its mechanical stability \cite{Alexander98}, leading to an abundance of low-frequency normal modes compared to crystalline structures. The properties of these modes have been extensively studied in disordered solids  \cite{Wyart05a, Lerner16}, with correlations drawn to solid rearrangements \cite{brito2009geometric, widmer2008irreversible}. They also have been proposed as elemental defects controlling the flow of disordered solids \cite{Manning2011}. It would be intriguing to investigate whether active particles distributed in specific lattice regions could selectively excite desired low-frequency mode intervals or unveil unpredictable features absent from the current work.

\section*{Conflicts of Interest}

There are no conflicts of interest to declare.

\begin{acknowledgments}
We thank Paul Baconnier for insightful discussions and the help provided on simulations. DL and CB thanks CNPq and CAPES for partially financing this study. We thank the supercomputing laboratory at New York University (NYU-HPC), where part of the simulations were run, for computer time.
\end{acknowledgments}


\bibliographystyle{apsrev4-1}
\bibliography{biblio}

\appendix

\section{Setting the disordered lattices}
\label{sec_appendix_disordered_lattice}
Disordered networks are created by first generating packings of soft discs at high density. To generate these packings, we employ a protocol used previously which is here summarized. We initiate by setting the density of particles $\rho = N_p/V =0.5$ (where $V$ represents the volume of the system and $N_p$ stands for the number of particles), followed by conducting a high-temperature equilibration (T = 1.0) of the system under the influence of the potential energy $U= \kappa/2\sum_{i,j} (r_{ij}-R_j-R_i)^2 H(R_j+R_i - r_{ij}) $, where $r_{ij}$ is the distance between the centers of particles $i$ and $j$, $R_j$ is the radius of the particle $j$ and $H(x)$ is the Heaviside function. Subsequently, we employ the FIRE algorithm  \cite{FIRE_PRL2006}  to minimize the potential energy U. Throughout this minimization process, we maintain a constant pressure of $p = 1.0$ using a Berendsen barostat \cite{Berendsen1984} with a time constant $\tau_{Ber} = 10.0$. The minimization procedure is stopped when the  interparticle force falls below $10^{-1}$. 
\begin{figure}[h!]
    \includegraphics[width=0.45\textwidth]{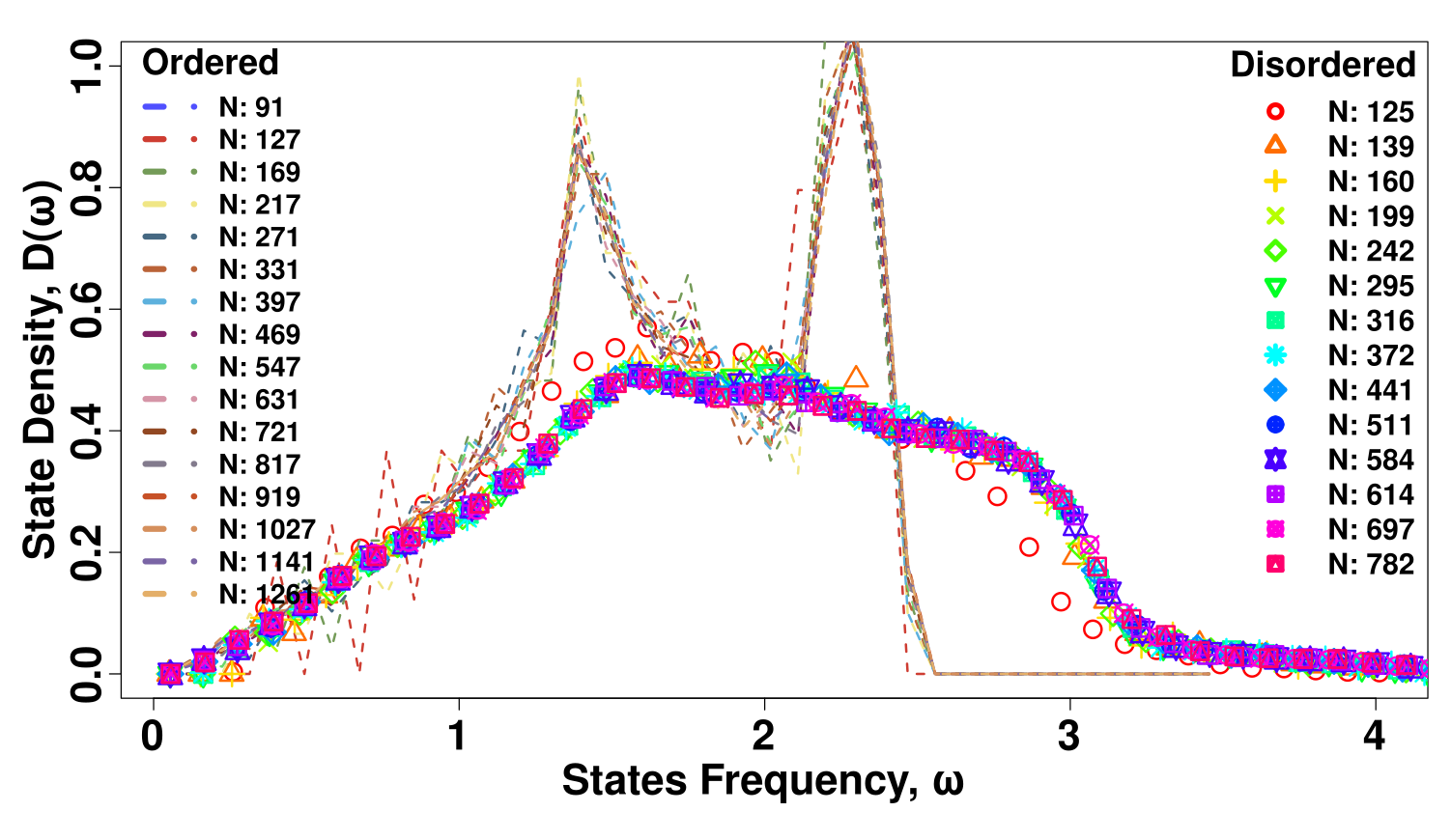}
    \caption{Density of states of the triangular lattice compared to the disordered case for several lattice sizes, showing that they are all statistically the same. }
    \label{Dw_ord_dis}
\end{figure}

The ordered and disordered networks can be compared in terms of the frequency distribution of their normal modes: Fig~\ref{Dw_ord_dis} shows the density of states $D(\omega)$, defined as the distribution of the frequencies $\omega$, that are the square root of the dynamical matrix ($\mathbb{M}$) eigenvalues, for both ordered and disordered networks. For small frequencies ($\omega < 1$), both types of lattices have similar distributions.

\begin{figure}[h]
    \includegraphics[width=0.45\textwidth]{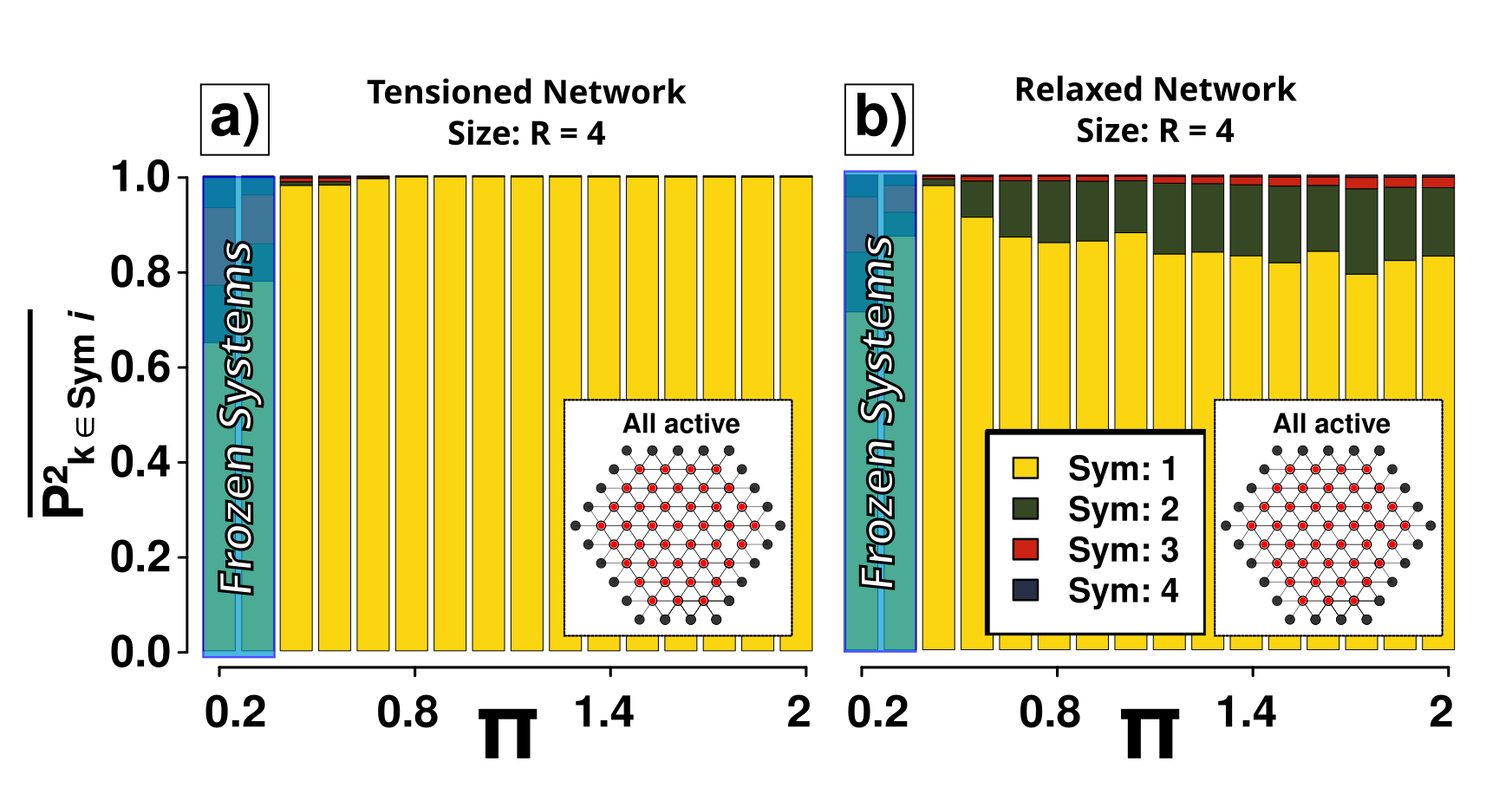}
        	\caption{Average projection of the dynamics on the normal modes of the lattice, divided into 4 rotational symmetries, as indicated in the legend, for a raging value of $\pi$.
          In {\bf (a)} the system is tensioned and in {\bf (b)} tension is zero.  Blue line indicates the percentage of moving systems for each $\pi$.  In both cases, the system size is $R=4$  and all the nodes of the lattice are full with active particles. Black points represent pinned edges, red points active nodes and open circles are empty nodes. Data shown here is an average over 100 run of dynamics starting with different polarities directions.}
    \label{P2k_R4}
\end{figure}

Additionally, the definition of the dynamical matrix does not account for any tension in the lattice. Figure \ref{P2k_R4} displays the average projection of the dynamics onto the normal modes of the lattice, categorized into four rotational symmetries. This is shown for the case where active particles occupy all the lattice nodes, as depicted in the figure's inset. The key difference between the two cases is the presence of tension in the lattice, which has been demonstrated to be a relevant control parameter in promoting different types of dynamics in active solids \cite{baconnier2023discontinuous}. In Figure \ref{P2k_R4}-a, the case with high tension is presented, showing that regardless of the value of $\Pi$, the dynamics project almost entirely onto symmetry type 1, recovering the results shown in \cite{baconnier2022selective}. When the tension is reduced, additional symmetries emerge in the dynamics, as shown in Figure \ref{P2k_R4}-b.

\section{Noise influence}
\label{noise}

To understand the implications of non-determinism on our system, we have also tested the convergence of the time series projection with the addition of noise. The noise is took into account on the Eq. \ref{eq_active2} (while Eq. \ref{eq_active1} remains unchanged), where it becomes:
\begin{eqnarray}
    \frac{d \bm{\hat n}_i}{dt} = \bm{\hat n}_i \times \bm{v}_i \times \bm{\hat n} + \sqrt{2D} \xi \bm{\hat n}^\perp_i,
\end{eqnarray}
for $D = \alpha \gamma^2 / \tau^2 \kappa^2$, where $\alpha$ is the noise amplitude, $\gamma$ is the drag coefficient, $\tau$ is the characteristic alignment time of the agent, $\kappa$ is the spring stiffness and $\xi$ is a unitary normal random variable.

\begin{figure}[h]
    \centering
    \includegraphics[width=\linewidth]{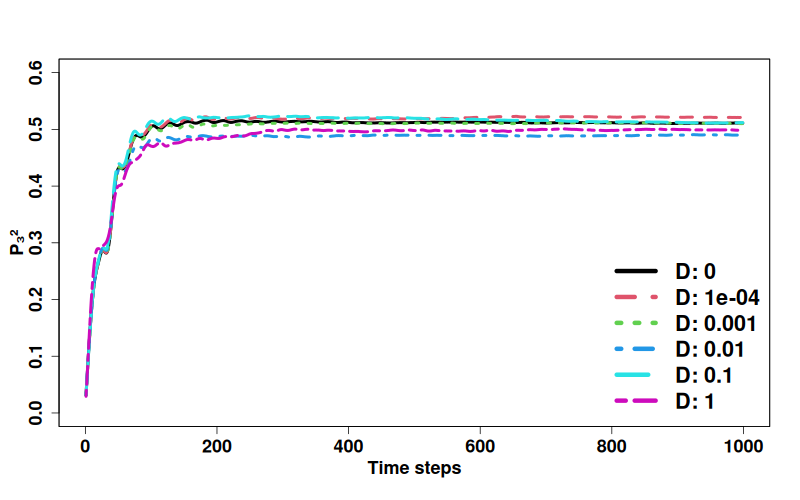}
    \caption{Influence of noise: average projections over the 3rd mode for 30 initial conditions, the system have size R = 7, and the active distribution (with $\phi_{dn} = 0.19$) is concentrated on $r=4$, for $\Pi = 1.3$.}
    \label{fig_time_serie}
\end{figure}

In Fig. \ref{fig_time_serie} it is shown the average values of ${{P_3^2}}$ where was considered a system with size R=7 and a density of active nodes of 24/127 – the same as in the Fig. \ref{fig_opt} of the manuscript – and distribute the 24 driven active nodes in the ring $r=4$ for 30 distinct initial conditions. In Fig. \ref{fig_time_serie} we observe that the  average projection of the dynamics on the third mode is robust to the addition of noise, even for noise of order one. 

\section{Robustness of effective temperature, $T_e$}
\label{det}

\begin{figure}[h]
\includegraphics[width=0.99\columnwidth]{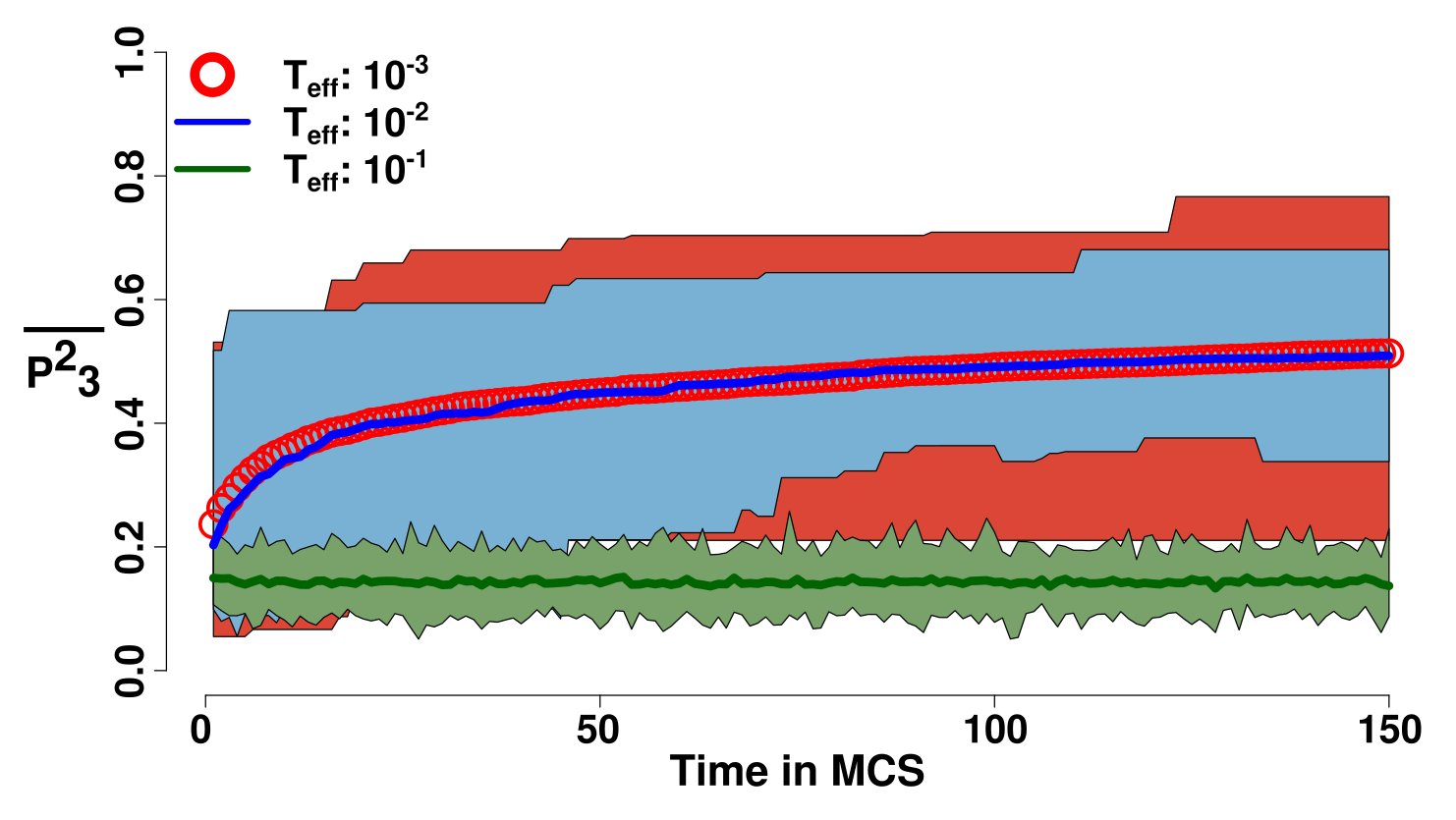}
\caption{Effective temperature, $T_e$, used on the optimization algorithm. On the graph with points and solid lines are represented the average of 100 optimizations and the shade areas are limited by the best (upper) and worst (lower) samples.}
\label{fig_variaTe} 
\end{figure}

To adjust the temperature like parameter $T_e$ used in the Monte Carlo rules for the optimization algorithm (Eq. \ref{eq_prob_change}), we explored three logarithmic scales of $T_e$. Fig.~\ref{fig_variaTe} displays the convergence of the algorithm for $T_e = 10^{-3}, T_e = 10^{-2}, T_e = 10^{-1}$. While too large $T_e=10^{-1}$ ruins the optimization dynamics, we don't see strong effect of $T_e$ in the range $[10^{-3}, 10^{-2}]$.

\section{Optimization algorithm time window}
\label{sec_time_window}

\begin{figure}[h]
    \centering
    \includegraphics[width=\linewidth]{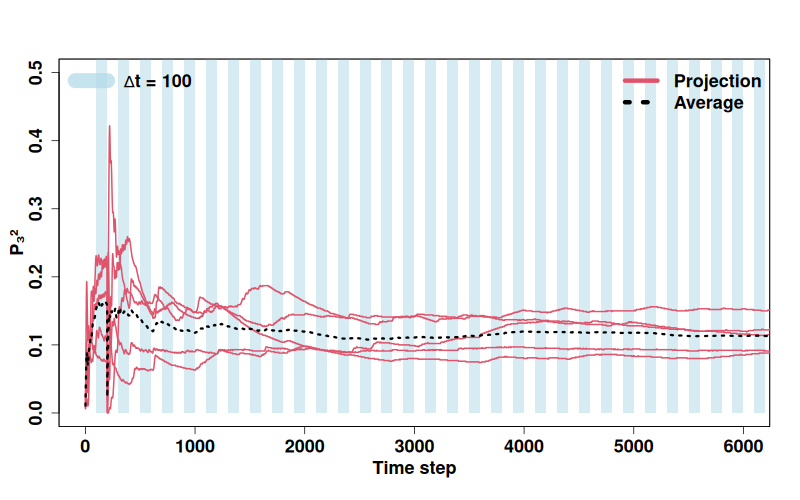}
    \caption{Optimization time series: projections over the 3rd mode for 5 initial conditions, the system have size R = 7, and random active distribution (with $\phi_{dn} = 24/127= 0.19$), for $\Pi = 1.3$. The blue intervals show the time window for measurements $\Delta t = 100$ and the white refers to intervals where no measurements are performed.}
    \label{fig_opt_time_serie}
\end{figure}

To test the robustness of the projection with respect to the time interval of measurement, we performed the following test. We considered a case with $R=7$ and a density of active nodes of $24/127$, where the 24 active particles are randomly distributed in the system. We attempted to change the location of an active particle every $2\Delta t = 200$  (molecular dynamics time steps) to optimize the projection of the dynamics onto mode 3 and summarized the results in Fig.~\ref{fig_opt_time_serie}.

Each red plot in this figure corresponds to the projection of the dynamics onto mode 3 for one initial condition (IC), while the black line represents the average over 5 ICs. The white regions indicate periods during which no measurements were taken, and the blue regions represent the measurement intervals of $\Delta t = 100$ time steps. Note that the dynamics quickly reach a steady state shortly after the simulation begins. Since only small changes are made every $2\Delta t = 200$ (with only one particle being moved while velocity and polarization parameters are maintained from one trial to the next), the dynamics converge rapidly, justifying the use of such measurement intervals.

\section{Modes of the ordered lattices with their class of symmetries}
\label{symmetries}


The triangular lattice with hexagonal boundaries respects symmetries over rotations of angle $\pi/3$, done by the operator $\Theta$, and reflections $\Sigma$ (e.g., across the axis $y = 0$). The eigenvectors of the dynamical matrix, $\mathbb{M}$, can be either direct eigenvectors of those symmetry operators or bases for inner spaces that respect those symmetries, with reflection eigenvalues given by $\sigma = \pm 1$ and rotation eigenvalues given by $\theta = \exp(i k \pi/3)$ for $k \in \{-2, \ldots, 3\}$.

The eigenvectors corresponding to the complex eigenvalues of the rotational operator are also complex and occur in degenerated pairs, $|\varphi_\pm \rangle$ relative to $\theta_\pm = \exp(\pm i n \pi/3)$ for some n. 
These paired modes can be combined into two real modes $|\varphi_l\rangle$ and $|\varphi_m \rangle$, which also have the same energy as $|\varphi_\pm \rangle$. Although $|\varphi_l \rangle$ and $|\varphi_m \rangle$ (eigenvectors of $\mathbb{M}$) are not eigenvectors of $\Theta$, the 2-dimensional space they span remains invariant under rotations of this kind. The effect of $\Theta$ on these modes is characterized by $\langle \varphi_l | \Theta | \varphi_l \rangle = \langle \varphi_m | \Theta | \varphi_m \rangle$, which corresponds to the real part of the eigenvalue of $|\varphi_\pm \rangle$. Therefore, the complete symmetry of a normal mode $|\varphi_k\rangle$ is determined by two real numbers:
\begin{gather*}
    \langle \varphi_k | \Theta | \varphi_k \rangle \in \{1, 1/2, -1/2, -1\} \ \mbox{and}\\
    \langle \varphi_k | \Sigma | \varphi_k \rangle \in \{1, -1\}.
\end{gather*}
In this work the normal modes of the dynamical matrix are characterized by the real part of the rotational symmetry eigenvalues they relate to. More precisely, the symmetries 1, 2, 3 and 4, used along the whole text, are related respectively to $\text{Real}(\theta) = 0.5, 1, -0.5 \text{ and } -1$. More details on the symmetry class derivation can be seen on the SI of \cite{baconnier2022selective}.

\begin{figure*}
\centering
    \includegraphics[width=1.9\columnwidth]{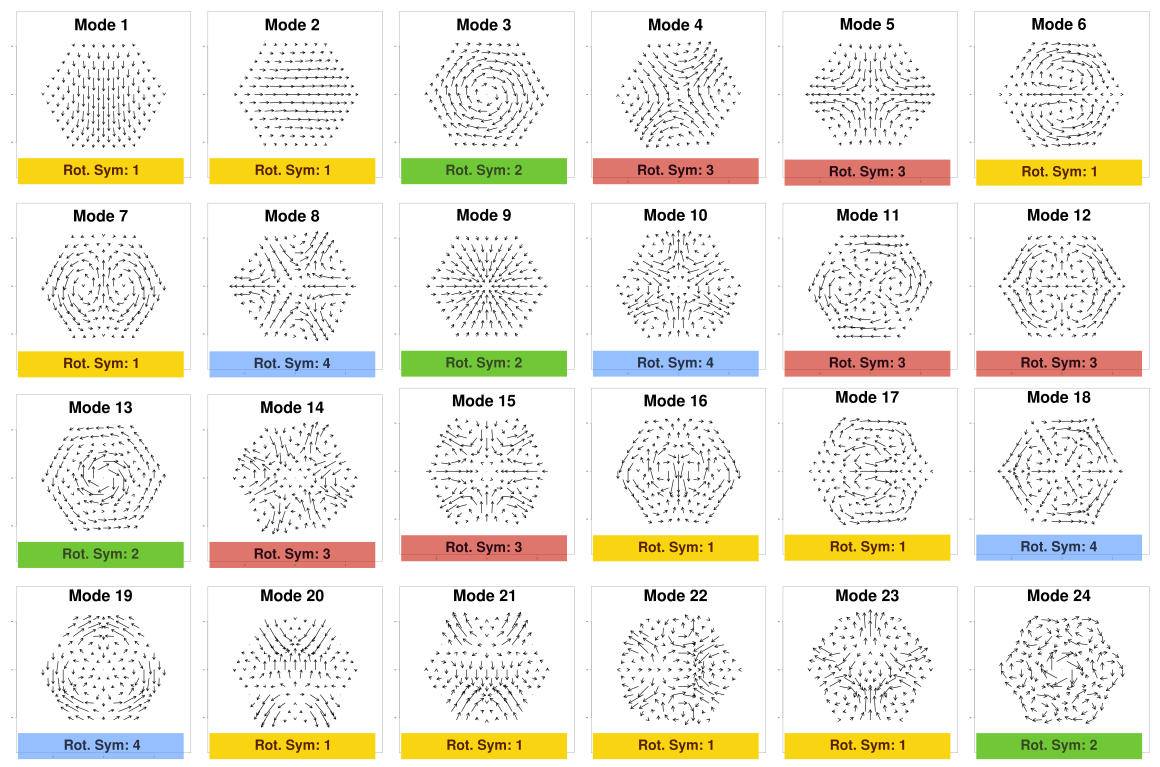}
    \caption{Normal modes displacement fields and rotational symmetries for modes from 1 to 24.}
    \label{fig_symmetry_modes} 
\end{figure*}

In Figure \ref{fig_symmetry_modes} are presented the first 24 normal modes for the ordered triangular lattice of size $R = 7$, the colors identify the class of rotational symmetry they follow.

\end{document}